\documentclass[journal=jpclcd,manuscript=letter]{achemso}

\usepackage[version=3]{mhchem} 
\usepackage[T1]{fontenc}       
\usepackage[usenames]{color}
\usepackage{graphicx}
\usepackage{float}
\author{Ana M. Valencia}
\affiliation{Humboldt-Universit\"{a}t zu Berlin, Physics Department and IRIS Adlershof, 12489 Berlin, Germany}
\alsoaffiliation{Carl von Ossietzky Universit\"at Oldenburg, Institute of Physics, 26129 Oldenburg, Germany}
\altaffiliation{Contributed equally to this work}
\author{Oleksandra Shargaieva}
\affiliation{Helmholtz-Zentrum Berlin für Materialien und Energie GmbH, Young Investigator Group 'Hybrid Materials Formation and Scaling', 12489 Berlin, Germany}
\altaffiliation{Contributed equally to this work}
\author{Richard Schier}
\affiliation{Humboldt-Universit\"{a}t zu Berlin, Physics Department and IRIS Adlershof, 12489 Berlin, Germany}
\author{Eva Unger}
\affiliation{Helmholtz-Zentrum Berlin für Materialien und Energie GmbH, Young Investigator Group 'Hybrid Materials Formation and Scaling', 12489 Berlin, Germany}
\author{Caterina Cocchi}
\affiliation{Humboldt-Universit\"{a}t zu Berlin, Physics Department and IRIS Adlershof, 12489 Berlin, Germany}
\alsoaffiliation{Carl von Ossietzky Universit\"at Oldenburg, Institute of Physics, 26129 Oldenburg, Germany}
\email{caterina.cocchi@uni-oldenburg.de}

\title{ Optical fingerprints of Polynuclear Complexes in Lead-Halide Perovskite Precursor Solutions}

\begin{document}
\newpage
\begin{center}
\section*{Abstract}
\end{center}
Solvent-solute interactions in precursor solutions of lead halide perovskites (LHP) critically impact the quality of solution-processed materials, as they lead to the formation of a variety of poly-iodoplumbates that act as building blocks for LHP. The formation of [PbI$_{2+n}$]$^{n-}$ complexes is often expected in diluted solutions  while coordination occurring at high concentrations is not well understood yet. In a combined \textit{ab initio} and experimental work, we demonstrate that the optical spectra of the quasi-one-dimensional iodoplumbates complexes
\ce{PbI2(DMSO)4}, \ce{Pb2I4(DMSO)6}, and \ce{Pb3I6(DMSO)8} formed in dimethyl sulfoxide solutions are compatible with the spectral fingerprints measured at high concentrations of lead iodide. This finding suggests that the formation of polynuclear lead-halide complexes should be accounted for in the interpretation of optical spectra of LHP precursor solutions.

\vspace{1.5cm}

\section*{Graphical TOC Entry}
\begin{center}
\includegraphics[height=5cm]{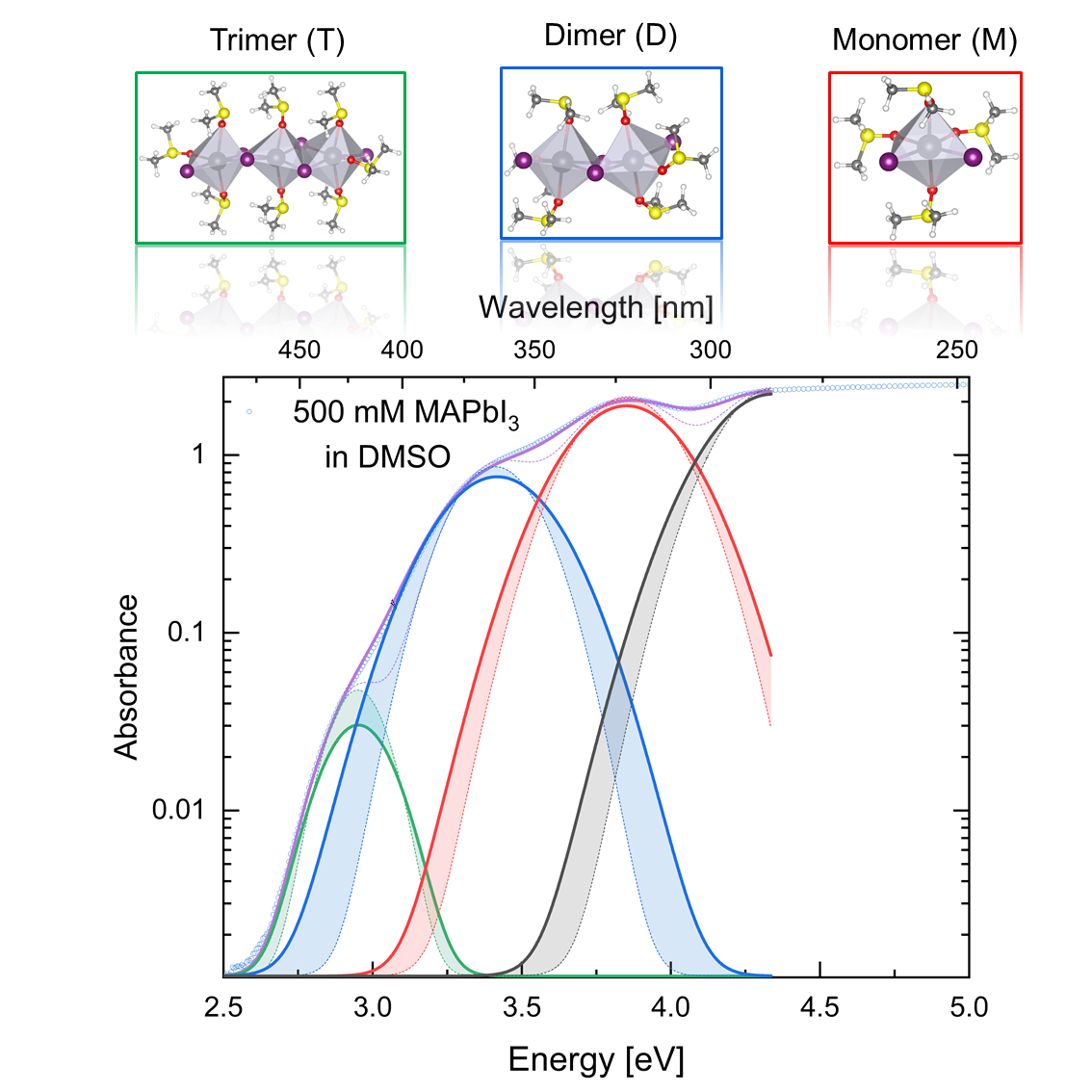}
\end{center}




\newpage

Lead halide perovskites (LHPs) are considered among the most promising solution-processed materials for the next generation of optoelectronic devices~\cite{snai13jpcl,jena+19cr}. The impressive progress in terms of device efficiency achieved within the last decade\cite{gree+20solar} has established the success of these systems and consolidated their popularity among solar cell materials. 
Similarly to other solution-processed materials, the quality of hybrid perovskites is dictated by solvent-solute interactions, by the structure of the intermediate solvate phases, and by the processing conditions of precursor solutions.~\cite{unge+14cm,zhou+15jpcl,chan+16acsami,ahla+19cm,jung+19csr} Therefore, open-ended questions regarding material formation range from the evolutionary process from solution complexes to thin films, the peculiar coordination of lead and iodine atoms, to the role of residual solvent molecules in altering the overall performance of LHPs.
Numerous experimental studies have reported the formation of coordination complexes in LHP precursor solutions, see \textit{e.g.}, Refs.~\citenum{rahi+16chpch,neno+16ees,yoon+16jpcl,radi+19acsaem}.

To date, the LHP solution chemistry and, more specifically, the formation of iodoplumbate complexes [PbI$_m$Sol$_n$]$_{2-m}$ (with $m = 0-4$ and $n = 0-6$; Sol indicates solvent molecules) has been demonstrated in diluted solutions with a concentration of the precursors in the range of 0.1 - 0.24 mmol/L, and a large excess of iodide ions (1:150)~\cite{rahi+16chpch,stam+15ees}. Recently, Radicchi \textit{et al.} rationalized with \textit{ab initio} calculations experimental optical spectra of such solutions in a number of common organic solvents and with different Pb:I ratios~\cite{radi+19acsaem}. In that work, the absorption spectrum of LHP precursor solution was interpreted through the red-shift of the absorption bands as the number of coordinated iodide ions (m) in [PbI$_m$Sol$_n$]$_{2-m}$ increases from 2 to 4. 

For thin-film processing, highly concentrated solutions in the range of 1 M are typically used. In such highly-concentrated solutions, the coordination complexes can be expected to differ significantly from mononuclear complexes observed in diluted solutions. In particular, interaction and agglomeration phenomena resulting in polynuclear complexes is expected. Sharenko \textit{et al.} suggested that the structure of the complexes formed in stoichiometric 1~M \ce{MAPbI3} in dimethylformamide is more similar to the structure of the crystalline intermediate phases formed during solidification than solution complexes~\cite{shar+17am}. Additionally, the formation of polynuclear iodoplumbates was hypothesized recently in highly concentrated solutions. Therefore, the formation of only mono-nuclear halido plumbates is insufficient for the rationalization of chemical interactions in the precursor solutions used for material deposition, as well as for understanding thin-film formation from solution. 
During crystallization of precursor solutions, the formation of intermediate perovskite phases with edge-sharing lead-iodide octahedra and inclusions of solvent molecules [\textit{e.g.}, (DMSO)$_2$(MA)$_2$Pb$_3$I$_8$] has been experimentally observed.~\cite{cao+16,petr+17jpcc,fate+18cm, shar+20ma}. Therefore, the agglomeration of iodoplumbates into polynuclear complexes is conceivable at a high concentration of precursors in the solution. These edge-sharing polynuclear complexes may act as predecessors for the crystalline intermediate phases during the early stages of crystallization.

The formation of quasi-one-dimensional polynuclear iodoplumbates has been previously observed in coordination compounds of PbI$_2$ with different organic ligands.~\cite{krau+01jcsdt,wu+09ccr,yu+14dt}. Typically, the increase of coordination in low-dimensional materials leads to a red-shift of the optical absorption due to the combination of enhanced screening and quantum confinement. Such a behavior has been discussed in relation to a number of diverse systems, including organic semiconductors,\cite{rent+99pccp,mall+07cp,cocc-drax15prb,vale-cocc19jpcc} graphene nanostructures,~\cite{baro+06nl,cocc+12jpcl,cocc+14jpca} and nanoparticles.~\cite{ougu+97prl,raty+03prl,bott-marq07prb}. Therefore, the formation of these polynuclear complexes can also affect the optical properties of LHP precursor solutions. 

To verify this hypothesis, we evaluate herein the impact  of polynuclear iodoplumbates formed in dimethyl sulfoxide (DMSO) solutions.
To this end, we investigate from first principles a set of neutral compounds with the chemical formula [PbI$_2$(DMSO)$_4$]$_N$ ($N= 1-3$). Through the calculation of the electronic structure and the optical absorption spectra from state-of-the-art many-body perturbation theory ($GW$ approximation and Bethe-Salpeter equation), we quantify the red-shift induced by the increasing Pb-I coordination, and provide an in-depth analysis of the spatial distribution of the lowest-energy excitations, which is useful to assess the impact of residual solvent molecules in LHPs.


%
\begin{figure}
\centering
\includegraphics[width=0.95\textwidth]{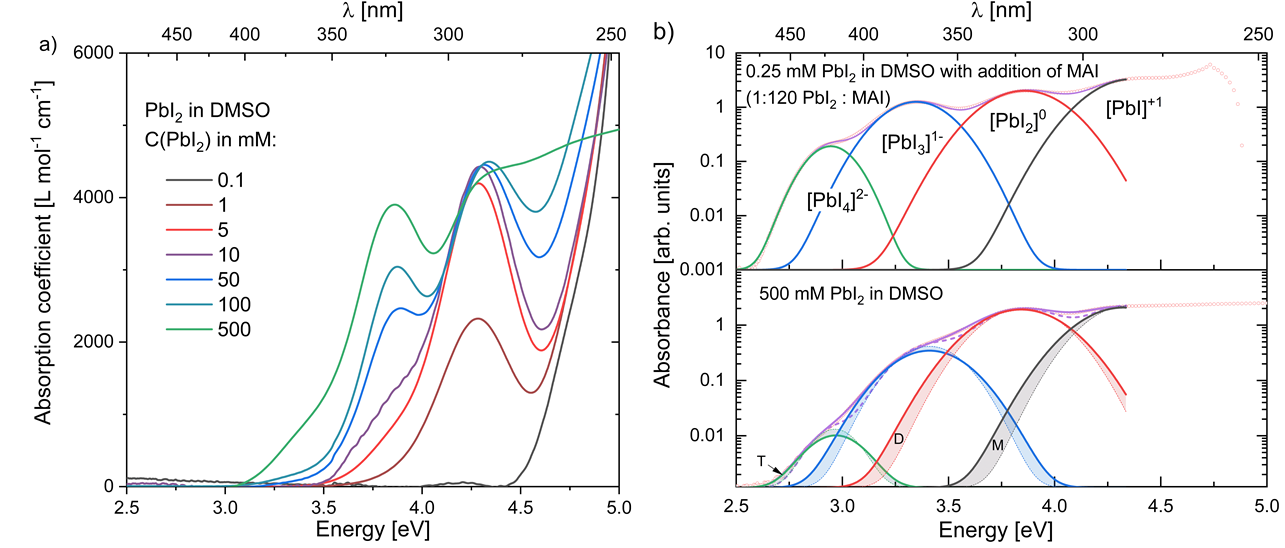}
\caption{a) Optical absorption spectra of PbI$_{2}$ in DMSO measured as a function of PbI$_{2}$ concentration, C(PbI$_{2}$). b) Fitted absorption spectra (red symbols) of 0.25~mM PbI$_2$ solution with addition of MAI (1:120 PbI$_2$:MAI) in DMSO (top) and 500~mM PbI$_2$ solution in DMSO (bottom). The fitted Gaussian functions indicate the positions of the absorption bands assigned to the iodoplumbate species [PbI]$^{+}$, [PbI$_2$]$^0$, [PbI$_3$]$^{-}$, and [PbI$_4$]$^{2-}$, respectively. Dashed lines indicate fitted Gaussian functions with peak width parameters kept as used for fitting spectrum of PbI2:MAI (1:120). Peak broadening associated with the formation of monomer (M), dimer (D), and trimer (T) structures is indicated by shaded areas.} 
   \label{fig:experimental-spectra}
\end{figure}

Experimental results motivating the \textit{ab initio} study presented herein are shown in Figure~\ref{fig:experimental-spectra}. The absorption spectra of lead iodide solutions in DMSO as a function of PbI$_2$ concentration are shown in Figure~\ref{fig:experimental-spectra}a). Low concentration (0.1-1~mM) lead iodide solutions exhibit absorption bands above 4.96~eV (\textit{i.e.}, below 250~nm) and 4.39~eV (282~nm), which correspond to the characteristic energies of dissociated PbI$_2$ species, [Pb(Sol)$_6$]$^{2+}$ and [PbI(Sol)$_5$]$^{+}$, respectively.~\cite{stam+15ees} Increasing the PbI$_2$ solution concentration to 5~mM leads to an increase of the absorption at 4.39~eV and to the appearance of a shoulder at 3.85~eV (325~nm). This can be attributed to the formation of [PbI$_2$(Sol)$_4$]$^0$. Further incrementing the PbI$_2$ concentration in solution results in the increase of the peak strength at 3.85~eV, and in the appearance of another feature/shoulder at 3.48~eV (357~nm) that can be attributed to [PbI$_3$(Sol)$_3$]$^{-}$. The solution with the highest concentration of PbI$_2$ in DMSO (500~mM) shows increased absorption below 3.1~eV (> 400~nm), in addition to the absorption band at 3.48~eV. 

Interestingly, the absorption bands observed in the solutions with a high concentration of PbI$_2$ in DMSO closely resemble the absorption spectrum of PbI$_2$ solutions with the addition of iodide ions. Figure~\ref{fig:experimental-spectra}b), top panel, shows the absorption spectra of PbI$_2$ with the addition of methylammonium iodide (MAI) (1:120 PbI$_2$:MAI), while the 500 mM PbI$_2$ solution in DMSO is shown in the bottom panel. The absorption spectra are fitted with Gaussian functions, indicated as shaded areas in Figure 1b). In the presence of an excess of MAI, the absorption bands at 3.48~eV and 2.98~eV were previously ascribed to the absorption of [PbI$_3$(Sol)$_3$]$^{-}$ and [PbI$_4$(Sol)$_2$]$^{2-}$ species.~\cite{stam+15ees} Using the same peak positions with a significant peak broadening, the Gaussian functions are fitted to the absorption spectrum of pure PbI$_2$. 

The position of the absorption bands of the concentrated PbI$_2$ solution indicates the presence of solution species that exhibit spectroscopic features similar to those previously correlated with the [PbI$_3$]$^{-}$ and [PbI$_4$]$^{2-}$. However, in the solutions of pure PbI$_2$, this observation cannot be simply rationalized by the formation of lead-halide solution complexes, as in the absence of MA$^+$ cations the charge neutrality of the solution species has to be preserved via interactions with other positively charged ions such as [Pb$^{2+}$Sol$_6$] and [PbI$_1$Sol$_5$]$^{+}$. 
Such interaction of iodoplumbate species with each other was previously attributed to the formation of polynuclear complexes in solution.~\cite{Shargaieva2020}
We interpret the broader absorption features around 3.48~eV of the highly concentrated PbI$_2$ solution with the interaction of lead halide species and the formation of polynuclear complexes. Solution of 500 mM \ce{MAPbI3} in DMSO exhibits similar broadening of the absorption bands as 500 mM solution of \ce{PbI2} (see Figure S1 in the Supporting Information).
Polynuclear complexes with two or three lead-halide-solvent units, such as [PbI$_2$(DMSO)$_N$]$_{2-3}$, hereafter referred to as dimers (D) and trimers (T), are critical initials steps in the formation of solid state perovskite semiconductors.
Considering the strong interaction of DMSO molecules with Pb$^{2+}$, we suggest that the interaction of the PbI$_2$ solution complexes leads to formation of edge-sharing (PbI$_2$)$_n$ polynuclear complexes, as suggested in previous work~\cite{Shargaieva2020}. 
In solvents with weaker coordination strength and consequently a larger amount of higher-coordinated charged lead-iodide solution complexes, we expect a higher likelihood for the formation of polynuclear complexes~\cite{hami+17acsel,stev+17cm,hami+20jpcc}. 

\begin{figure}
\centering
\includegraphics[width=0.48\textwidth]{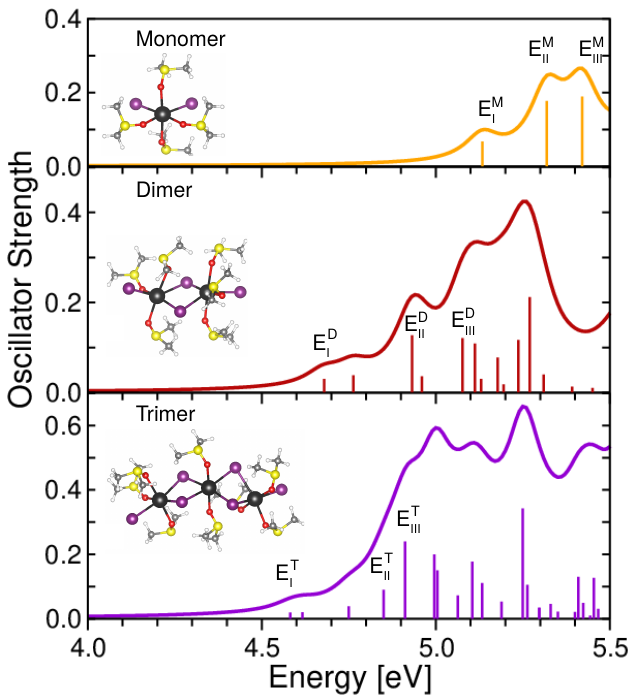}
\caption{Optical absorption spectra computed for the monomer, dimer, and trimer structures PbI$_{2}$(DMSO)$_{4}$, Pb$_{2}$I$_{4}$(DMSO)$_{6}$, and Pb$_{3}$I$_{6}$(DMSO)$_{8}$ respectively, shown in the inset. Vertical bars indicate the energy and oscillator strength of the calculated excitations.
A Lorentzian broadening of 500~meV is applied to all spectra.}
\label{fig:spectra}
\end{figure}

The iodoplumbate complexes in DMSO solution corresponding to monomer (M), dimer (D), and trimer (T) structures introduced above are modeled by the chemical formula [PbI$_{2}$(DMSO)$_{4}$]$_N$, where $N$ runs from 1 to 3, respectively (see Figure~\ref{fig:spectra}, insets).
Each PbI$_{2}$ unit is coordinated with 4 DMSO molecules to ensure charge neutrality.
The resulting structures are obtained via force minimization in the framework of density-functional theory (DFT)~\cite{hohe-kohn64pr,kohn-sham65pr} (further details are provided below, in the Section Theoretical Methods and Computational Details). 
Interestingly, the dimer and trimer exhibit the polyhedral coordination of the polynuclear iodoplumbate complexes reported in Ref.~\citenum{krau+01jcsdt}.
A close inspection on the optimized model structure in the inset of Figure~\ref{fig:spectra} reveals their quite distorted backbone (see Supporting Information, Figure~S2 and related discussion), which was not reported in previous works~\cite{krau+01jcsdt,wu+09ccr,yu+14dt}.
We ascribe this behavior to our choice of simulating the iodoplumbate complexes as isolated compounds interacting atomistically with DMSO solvent molecules linked to the Pb atoms through the available oxygen lone pair. 
Additionally, we considered these compounds \textit{in vacuo}, in order to focus on the \textit{explicit} quantum-mechanical interactions between solute and solvent without additional contributions from implicit media which would further enhance the electronic screening.
As a consequence of these choices, we anticipate a systematic overestimation on the order of 1~eV of the excitation energies with respect to the experimental ones reported in Figure~\ref{fig:experimental-spectra}.
However, we do not expect the physical picture provided by our first-principles results to be altered by this quantitative discrepancy.

The optical absorption spectra computed for the M, D, and T model compounds (see Figure \ref{fig:spectra}) show a significant red-shift at increasing coordination length of the \ce{PbI2} units. 
All three systems exhibit a smooth absorption onset characterized by a bright but relatively weak lowest-energy excitation, labeled E$_I$.
In the monomer, E$_I^{M}$ is found at 5.13~eV, in the dimer E$_I^{D}$ is at 4.67~eV, and in the trimer E$_I^{T}$ appears at 4.58~eV. 
The oscillator strength (OS) calculated for E$_I$ decreases from the monomer to the trimer (see Figure \ref{fig:spectra} and Table~S1 in the Supporting Information for further details).
These results clearly indicate that the energy of the first excitation decreases at increasing complex length, hinting at a similar behavior as the inverse power law $1/l$ ($l$ is the \textit{effective length} of the system). 
The latter is a common behavior of quantum-confined nanosystems observed also, for example, in the context of carbon-based nanostructures~\cite{han+07prl,baro+06nl,cocc+12jpcl,cocc+14jpca}. 
This given, the overall intensity associated to the onset appears similar in all spectra, due to the presence of additional excitations in the spectra of the dimer and trimer, which are energetically to the first one. 

Beyond the onset, the absorption increases steeply in all spectra (see Figure \ref{fig:spectra}).
In the spectrum of the monomer, only two excitations appear beyond the first one, which we label E$_{II}$ and E$_{III}$.
In the spectra of dimer and trimer, many more excited states and absorption maxima characterize the energy region between 4.8~eV and 5.5~eV. 
From the analysis of these excitations in terms of single-particle transitions, we identify two maxima that can be associated with E$_{II}$ and E$_{III}$.
From these results, we can also calculate hole and electron densities that quantify their spatial distribution across solute and solvent~\cite{cocc+11jpcl,vale-cocc19jpcc,vale+2020pccp}:
\begin{equation}
\rho_{h}^{\lambda} (\textbf{r})=\sum_{\alpha \beta} A_{\alpha \beta}^{\lambda} \rvert \phi_{\alpha}(\textbf{r})\lvert^{2}
\label{eq:h}
\end{equation}
and
\begin{equation}
\rho_{e}^{\lambda} (\textbf{r})=\sum_{\alpha \beta} A_{\alpha \beta}^{\lambda} \rvert \phi_{\beta}(\textbf{r})\lvert^{2},
\label{eq:e}
\end{equation}
respectively, where $\phi_{\alpha}$ and $\phi_{\beta}$ are the occupied and unoccupied molecular orbitals included in the transition space, and $A_{\alpha \beta}^{\lambda}$ are the absolute squares of the normalized BSE coefficients, which act as weighting factors of $\phi_{\alpha}$ and $\phi_{\beta}$ for the $\lambda$-th excitation. 

\begin{figure} 
\centering
\includegraphics[width=1.0\textwidth]{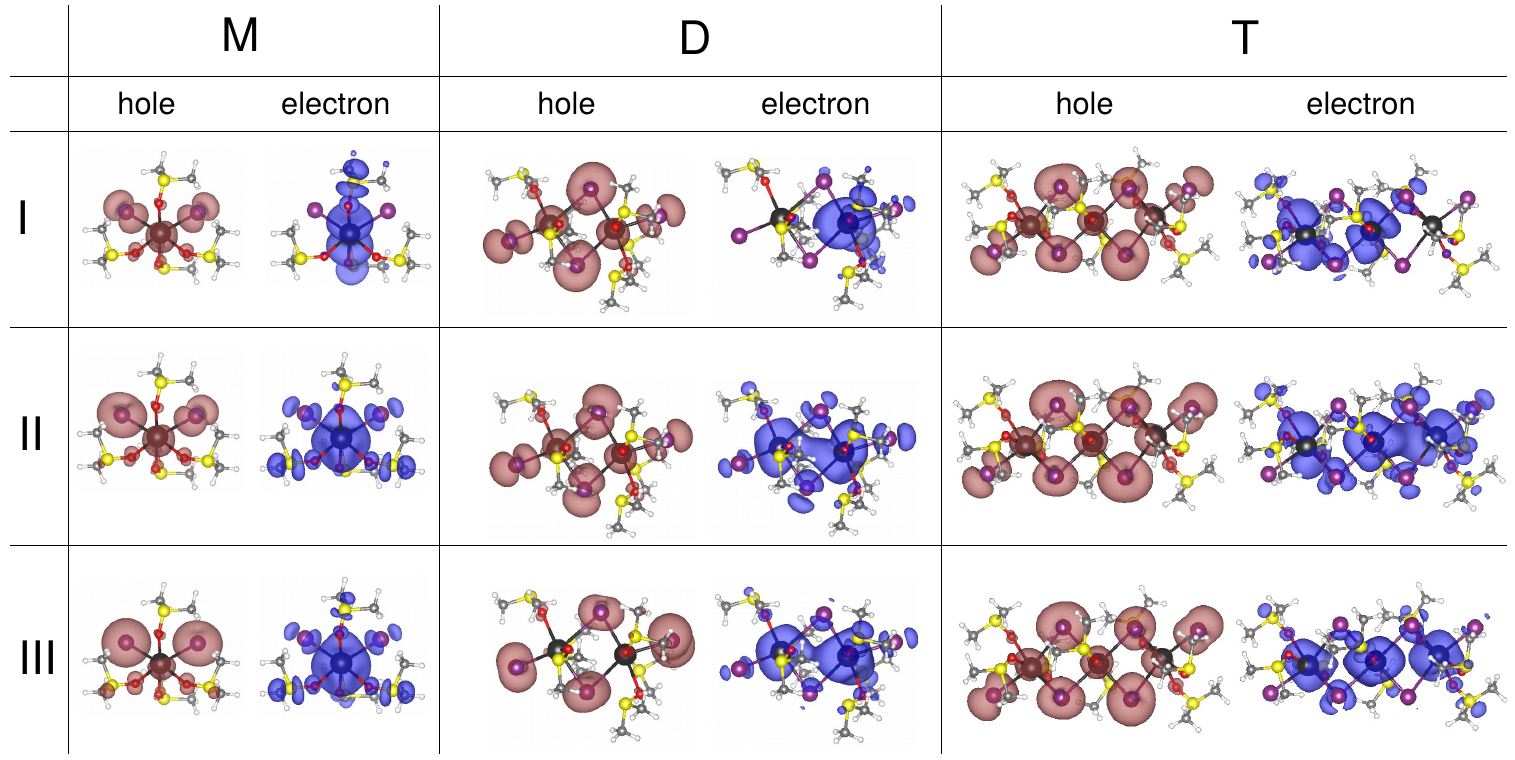}
\caption{Hole and electron densities calculated for E$_{I}$, E$_{II}$ and E$_{III}$ in the monomer (M), dimer (D), and trimer (T). Isosurfaces are plotted with a cutoff of 0.002~\AA{}$^{-3}$. }
\label{fig:h-e-densities}
\end{figure}

The results summarized in Figure~\ref{fig:h-e-densities} offer important insight for the overall understanding of the optical excitations in iodoplumbate precursors of LHPs in solution.
First of all, it is evident that upon photoexcitation both the electron and the hole are essentially localized on the Pb-I backbone with very little contribution from the DMSO solvent molecules. 
This is a very relevant result, as it is proves that regardless of the formed complex structure  between solute and solvent molecules, the latter essentially do not participate in the optical absorption. 
From this finding we speculate that the solvent molecules coordinating solution-complexes and incorporated in solvate-intermediate phases of LHP thin-films do not themselves affect the absorption spectrum of the material and that the absorption features stem from the lead iodide bonding interaction. Solvent-coordination does, however, critically affect the nature and structure of solution building blocks and their polynuclear assemblies.
The second important evidence emerging from the results in Figure~\ref{fig:h-e-densities} is the increasing delocalization of the electron and hole densities across the molecular backbone as the latter increases in size. 
In the monomer, a non-negligible portion of the electron density and, to a lower extent, also of the hole density are spilled-over onto the oxygen atom (E$^{M}_I$) and also on the sulphur atoms (E$^{M}_{II}$ and E$^{M}_{III}$). 
This characteristic is closely connected with the character and the spatial distribution of the molecular orbitals contributing to these excitations (see Supporting Information, Table~S1 and Figure~S3). 

In the monomer, the relatively small spatial extension of the molecular backbone enhances the hybridization between solute and solvent also in the orbitals closest to the fundamental gap. 
The first excitation, E$^{M}_{I}$, stems entirely from the transition from the highest-occupied molecular orbital (HOMO) to the lowest-unoccupied orbital (LUMO). 
Also E$^{M}_{II}$ corresponds to a transition from the HOMO mainly targeting the LUMO+1. 
Finally, the highest-energy of the examined excitations, E$^{M}_{III}$, is formed by a mixture of different transitions, including the one from the HOMO to the LUMO+2, as well as from the HOMO-1 to the LUMO+1.
Because of their similar composition in terms of single-particle transitions, E$^{M}_{II}$ and E$^{M}_{III}$ exhibit very similar electron and hole densities.
In the dimer, the first excitation, E$^{D}_{I}$, also stems from the HOMO$\rightarrow$LUMO transition, although the contribution from HOMO-1$\rightarrow$LUMO is non-negligible (see Table~S1). E$^{D}_{II}$ exhibit a similar composition as in the monomer, while E$^{D}_{III}$ has a strongly mixed character, given by a number of transitions between the HOMO and the first few lowest-unoccupied orbitals. Notably, all the virtual states involved are largely delocalized along the iodoplumbate backbone (see Figure~S3), as testified by the corresponding electron distribution in Figure~\ref{fig:h-e-densities}.
The longer extension of the trimer enhances the number of molecular orbitals contributing to the low-energy excitations.
In E$^{T}_{I}$, contributions from the HOMO-1$\rightarrow$LUMO and the HOMO$\rightarrow$LUMO transitions can still be identified, while both E$^{T}_{II}$ and E$^{T}_{III}$ stem from a huge number of transitions reported in Table~S2.

It is worth noting that the mixing of single-particle transitions is a signature of correlation effects that are quantitatively captured by the adopted \textit{ab initio} many-body methodology. 
The spatial distribution of the electron and hole densities, as analyzed above, is consistent with the relative OS of the corresponding excitations. 
In particular, the large overlap between electron and hole, combined with their extension along almost the entire length of the compounds is responsible for the enhanced absorption at increasing number of iodoplumbate units.
Even larger yield is expected in extensively coordinated LHP films.

After the analysis of the \textit{ab initio} results, we can go back to the experimental spectra shown in Figure~\ref{fig:experimental-spectra} and critically discuss our hypothesis, that the observed red-shift of the peaks at increasing solute concentration is related to the formation of polynuclear iodoplumbate complexes. 
Similarly to the experimental data, the spectra computed for the three modeled structures [PbI$_{2}$DMSO$_{4}$]$_{N}$ indeed exhibit a spectral shift to lower energies, as $N$ increases from 1 to 3 (see Figure~\ref{fig:spectra}).   
These systems exhibit similar optical properties to previously reported polynuclear plumbate complexes formed in LHP precursor solutions (Figure~\ref{fig:experimental-spectra}).~\cite{rahi+16chpch,yoon+16jpcl,radi+19acsaem} This observation is consistent with the fact that calculated hole and electron densities of the iodoplumbate complexes are mainly localized on the Pb and I atoms with only negligible contributions arising from solvent molecules. Thus, our results confirm the hypothesis that the absorption bands observed at high concentration of lead iodide solutions indeed correspond to the presence of polynuclear iodoplumbate complexes. It is important to note that most of the previously reported studies used a large excess of I$^-$ (\textit{e.g.}, 1:150 ratio of PbI$_2$:MAI) to generate spectroscopically detectable amounts of [PbI$_4$Sol$_2$]$^{2-}$ complexes, whilst 500 mM solution PbI$_2$ has already showed a considerable amount of trimer species.~\cite{rahi+16chpch,yoon+16jpcl,radi+19acsaem} Hence, in LHP precursor solutions with 1:1 ratio of PbI$_2$:MAI, the formation of polynuclear iodoplumbates becomes quite likely. 
This suggests that absorption spectra of LHP precursor solutions at high concentration cannot be  interpreted solely in terms of species with chemical formula [PbI$_{2+n}$Sol$_m$]$^{n-}$: The presence of polynuclear iodoplumbates should be accounted for too.

In summary, we were able to provide an interpretation of experimental absorption spectra of LHP precursors in solutions from \textit{ab initio} many-body simulations of polynuclear iodoplumbate complexes of increasing length, atomistically interacting with solvent molecules.
Our findings confirm that such species exist in solution and have similar optical properties than plumbate complexes that are usually considered as LHP solution complexes.
Considering the strong interaction of DMSO molecules with Pb, we suggest that solvent with weaker coordination strength should enhance the formation of polynuclear complexes.
Further studies are required to establish possible connections between such complexes and intermediate structures formed during the  crystallization of LHP thin-films.

\section*{Theoretical and Computational Methods}
Ground- and excited-state properties of the systems considered in this work are calculated from DFT and MBPT.
The [PbI$_{2}$(DMSO)$_{4}$]$_N$ complexes with $N=1-3$ are constructed assuming Pb-I units on the same plane and the optimized DMSO molecules occupying the remaining coordination sites.
In the intial geometries, Pb-I and Pb-O bonds are set equal to 2.5~\AA{} and then relaxed via force minimization (threshold 10$^{-3}$~eV/\AA{}) in the framework of DFT.
These calculations are carried out with the all-electron package FHI-aims~\cite{blum+09cpc} adopting tight integration grids and TIER2 basis sets~\cite{havu+09jcp} and the Perdew-Burke-Ernzerhof~\cite{perd+96prl} paramterization of the generalized-gradient approximation for the exchange-correlation functional.
Van der Waals interactions are explicitly accounted through the Tkatchenko-Scheffler scheme~\cite{tkat-sche09prl}. 

MBPT calculations are performed using the MOLGW code~\cite{brun+16cpc}. 
Gaussian-type cc-pVDZ basis sets for the light atoms and effective core potential (ECP) cc-pVDZ-PP basis sets for Pb and I atoms~\cite{brun12jcp} are used. Spin-orbit coupling is included in the ECP at the scalar relativistic level. In these calculations, the resolution-of-identity approximation is employed~\cite{weig+02}. 
The range-separated hybrid functional CAM-B3LYP~\cite{yana+04cpl} is adopted in the DFT calculations acting as starting point for the subsequent application of $G_0W_0$ and BSE. 

The total number of occupied and unoccupied states in the transition space adopted for the solution of the BSE, solved within the Tamm-Dancoff approximation~\cite{rang+17jcp}, is determined by the number of basis functions in the DFT step. In the monomer, this amounts to a total of 429 KS states, including 120 occupied and 309 virtual orbitals. For the dimer 198 occupied and 480 virtual orbitals are used, while in the trimer, 276 occupied and 651 virtual orbitals are considered. Within the frozen-core approximation adopted in the solution of the BSE, the total number of occupied states is effectively reduced to 88 in the monomer, 150 in the dimer, and 212 in the trimer.

\section*{Experimental Methods}
Solutions of PbI$_2$ were prepared by dissolving an appropriate amount of PbI$_2$ powder (99.999\% trace metal) in anhydrous DMSO in N$_2$ atmosphere. The solutions were shaken at 60$^{\circ}$~C for 12 h and cooled down to room temperature before use. Absorption measurements of 500-10 mM PbI$_2$ solutions were conducted in 10 $\mu$L cuvette. The optical path was increased to 0.1~cm for absorption measurements of solutions with concentrations lower than 10~mM. Solutions of PbI$_2$ with addition of methylammonium iodide was measured in a 10~$\mu$L cuvette.

\begin{acknowledgement}

This work was supported by the German Research Foundation (DFG), Priority Programm SPP 2196 - Project number 424394788.
C.C. acknowledges funding from the German Federal Ministry of Education and Research (Professorinnenprogramm III) as well as from the State of Lower Saxony (Professorinnen für Niedersachsen).
Computational resources were provided by the North-German Supercomputing Alliance (HLRN), project bep00076.

\end{acknowledgement}


\begin{suppinfo}
Additional details about experiments and calculations are reported in the Supporting Information. 
\end{suppinfo}

\newpage

\providecommand{\latin}[1]{#1}
\makeatletter
\providecommand{\doi}
  {\begingroup\let\do\@makeother\dospecials
  \catcode`\{=1 \catcode`\}=2 \doi@aux}
\providecommand{\doi@aux}[1]{\endgroup\texttt{#1}}
\makeatother
\providecommand*\mcitethebibliography{\thebibliography}
\csname @ifundefined\endcsname{endmcitethebibliography}
  {\let\endmcitethebibliography\endthebibliography}{}

\end{document}